\documentclass{webofc}\usepackage[varg]{txfonts}\usepackage{hyperref}\usepackage{url}
\usepackage{epsfig,graphics,amssymb,amsmath,mathrsfs,color,colortbl,bm,booktabs,dcolumn}
\hypersetup{colorlinks=true,citecolor=blue,urlcolor=blue,linkcolor=blue}
\begin{document}
\title{Various Illustrative Brief Glances at the\\Flavour-Cryptoexotic Tetraquark States}
\author{\firstname{Wolfgang} \lastname{Lucha}\inst{1}\fnsep\thanks{\email{Wolfgang.Lucha@oeaw.ac.at}}}
\institute{Institute for High Energy Physics, Austrian Academy of Sciences, Nikolsdorfergasse 18,\\A-1050 Vienna, Austria}
\abstract{Constituting the largest subset of multiquark states so far observed by experiment, tetraquark mesons carrying overall quark flavour quantum numbers identical to those of the (conventional) quark–antiquark mesons necessitate their particularly deliberate phenomenological analyses. A collection of (more or less recent) insights specific to the latter subclass of exotic hadrons bears the~promise to enable even significant improvement in the understanding of these hadrons by theoretical approaches such as, among others, the framework of QCD sum~rules.}
\maketitle

\section{Cryptoexotic tetraquarks as nature's favourite multiquark hadrons}\label{crex}Colour confinement restricts all (observable) bound states of dynamical degrees of freedom of quantum chromodynamics to colour singlets under the action of its gauge group SU(3). These states comprise not only all conventional hadrons --- quark--antiquark mesons and three-quark or three-antiquark baryons --- but also several variants of exotic hadrons: multiquarks (such as \textsl{mesonic} tetraquarks, hexaquarks, \textsl{etc.}\ and \textsl{baryonic} pentaquarks, heptaquarks, \textsl{etc.}) as~well as hybrids composed of quarks and gluons, and purely gluonic bound states denoted as glueballs.

At least at present, the most densely populated \textsl{sub}-category of experimentally established multiquark exotic hadrons is the set of tetraquark mesons --- constituted by the bound states of two quarks $q$ and two antiquarks $\overline q$, with flavour quantum numbers $a,b,c,d\in\{u,d,s,c,b\}$ and masses $m_a,m_b,m_c,m_d$ --- of the \textsl{flavour-cryptoexotic} sort; for the latter defining characteristic, this subset of exotic hadrons certainly deserves particularly enhanced theoretical attention \cite{WL:MT2}:\begin{quote}\textbf{Definition:} A \textsl{multiquark} hadron state is termed \textsl{``flavour-cryptoexotic''} if it does not feature more \textsl{open} quark flavours than a respective \textsl{conventional} hadron state.\end{quote}By comprising \textsl{at least} one quark--antiquark pair of one and the same flavour, any tetraquark $T$ of flavour-cryptoexotic type exhibits at most two quark flavours, with immediate implications:\begin{equation}T=[\overline q_a\,q_b\,\overline q_b\,q_c]\ ,\qquad a,b,c\in\{u,d,s,c,b\}\ .\label{fce}\end{equation}\begin{itemize}\item It may therefore undergo \textsl{mixing} with all the conventional-meson {open-flavour} counterparts\begin{equation}M_{\overline ac}=[\overline q_a\,q_c]\ ,\qquad a,c\in\{u,d,s,c,b,t(,\dots?)\}\ .\label{cm}\end{equation}\item Its \textsl{multiquark} identity cannot be unambiguously told just from its number of \textsl{open} flavours.\end{itemize}Extracted from Table~1 of Ref.~\cite{LMSurp}, Table~\ref{TC} recalls \cite{WL:MT2} all possible quark-flavour configurations as well as all conceivable antiquark--quark pairings of the flavour-cryptoexotic tetraquarks (\ref{fce}).

\begin{table}[ht]\caption{Breakdown \cite{LMSurp} of the quark-flavour content, $a\ne b\ne c$, of the \textsl{flavour-cryptoexotic} tetraquarks (\ref{fce}), in terms of involved and \textsl{open} (\textsl{viz.}, not counterbalanced by respective \textsl{antiflavours}) quark flavours.\label{TC}} \begin{center}\begin{tabular}{ccc}\toprule\textbf{Number of Different Flavours}&\textbf{$\hspace{1.08522ex}$Quark Composition$\hspace{1.08522ex}$}&\textbf{Number of Open Flavours}\\\midrule 3&$\overline q_a\,q_b\ \,\overline q_b\,q_c$&2\\&$\overline q_a\,q_b\ \,\overline q_c\,q_c$&2\\\midrule 2&$\overline q_a\,q_a\ \,\overline q_a\,q_b$&2\\&$\overline q_a\,q_a\ \,\overline q_b\,q_a$&2\\&$\overline q_a\,q_b\ \,\overline q_b\,q_a$&0\\&$\overline q_a\,q_a\ \,\overline q_b\,q_b$&0\\\midrule 1&$\overline q_a\,q_a\ \,\overline q_a\,q_a$&0\\\bottomrule\end{tabular}\end{center}\end{table}

\section{Relevance of the tetraquark contributions to four-point correlators}\label{4.cor}A promising and rather convenient starting point for theoretical explorations of any tetraquark meson is (hardly surprisingly) offered by the time-ordered (T) four-point correlation functions\begin{equation}\left\langle{\rm T}\!\left(j(y)\,j(y')\,j^\dag(x)\,j^\dag(x')\right)\right\rangle\label{4}\end{equation}of appropriately defined local quark--antiquark bilinear operators (sometimes named currents)\begin{equation*}j_{\overline ab}(x)\equiv\overline q_a(x)\,q_b(x)\ ,\end{equation*}wherein all reference to the (for the following irrelevant) spin or parity degrees of freedom has been neglected. The correlation functions (\ref{4}) govern, among others, all amplitudes describing scatterings of two \textsl{conventional} mesons (\ref{cm}), of momenta $p_1$ and $p_2$, into two \textsl{ordinary} mesons. In these correlation functions, tetraquark mesons manifest in form of poles in the Mandelstam variable $s\equiv(p_1+p_2)^2$. Accordingly, for identifying uniquely \textsl{all} genuinely ``tetraquark-phile'' \cite{TQPa,TQPb} contributions to any correlation function (\ref{4}), a selective \textsl{criterion} \cite{TQC} has been proposed:\begin{quote}\textbf{Proposition:} A truly tetraquark-phile contribution to any correlation function (\ref{4}) depends on $s$ in a non-polynomial fashion and exhibits a four-intermediate-quark \textsl{branch cut} that starts at the associated \textsl{branch point} at $\hat s\equiv(m_a+m_b+m_c+ m_d)^2$~\cite{TQC}.\end{quote}The cut's existence or nonexistence can be established by exploiting the Landau equations~\cite{LDL}.

To illustrate the basic ideas underlying this concept, combinatorial simplicity suggests, for definiteness, to focus to that species of tetraquarks among all the configurations categorized in Table~\ref{TC} that involve precisely three different quark flavours \cite{CEa,CEb}, \textsl{viz.}, to the tetraquark~subset\begin{equation}T=[\overline q_a\,q_b\,\overline q_b\,q_c]\ ,\qquad a,b,c\in\{u,d,s,c,b\}\ ,\qquad a\ne b\ne c\ .\label{3F}\end{equation}In form of intermediate-state poles, these tetraquarks may show up in the correlation functions\begin{eqnarray}&\left\langle{\rm T}\!\left(j_{\overline ab}(y)\,j_{\overline bc}(y')\,j^\dag_{\overline ab}(x)\,j^\dag_{\overline bc}(x')\right)\right\rangle,\qquad\left\langle{\rm T}\!\left(j_{\overline ac}(y)\,j_{\overline bb}(y')\,j^\dag_{\overline ac}(x)\,j^\dag_{\overline bb}(x')\right)\right\rangle,&\label{cfp}\\&\left\langle{\rm T}\!\left(j_{\overline ab}(y)\,j_{\overline bc}(y')\,j^\dag_{\overline ac}(x)\,j^\dag_{\overline bb}(x')\right)\right\rangle&\label{cfr}\end{eqnarray}of either identical (\ref{cfp}) or different (\ref{cfr}) initial and final distributions of their three quark flavours.

\begin{figure}[htb]\centering\includegraphics[width=8.0866cm,clip]{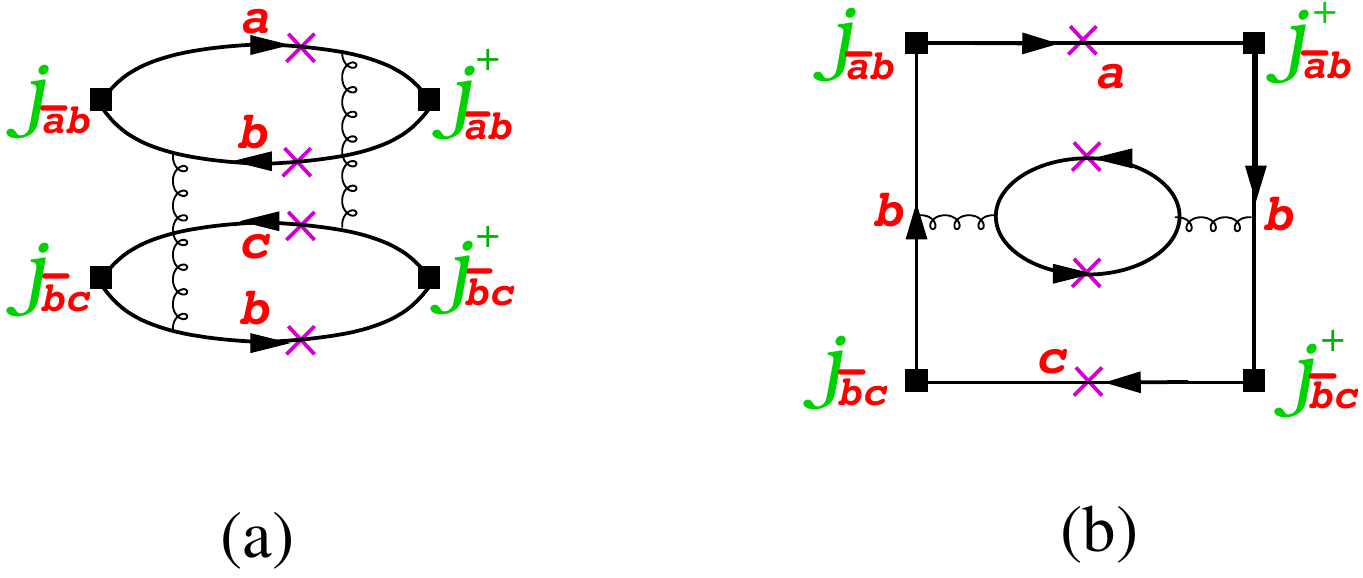}\\[1ex]\includegraphics[width=8.0866cm,clip]{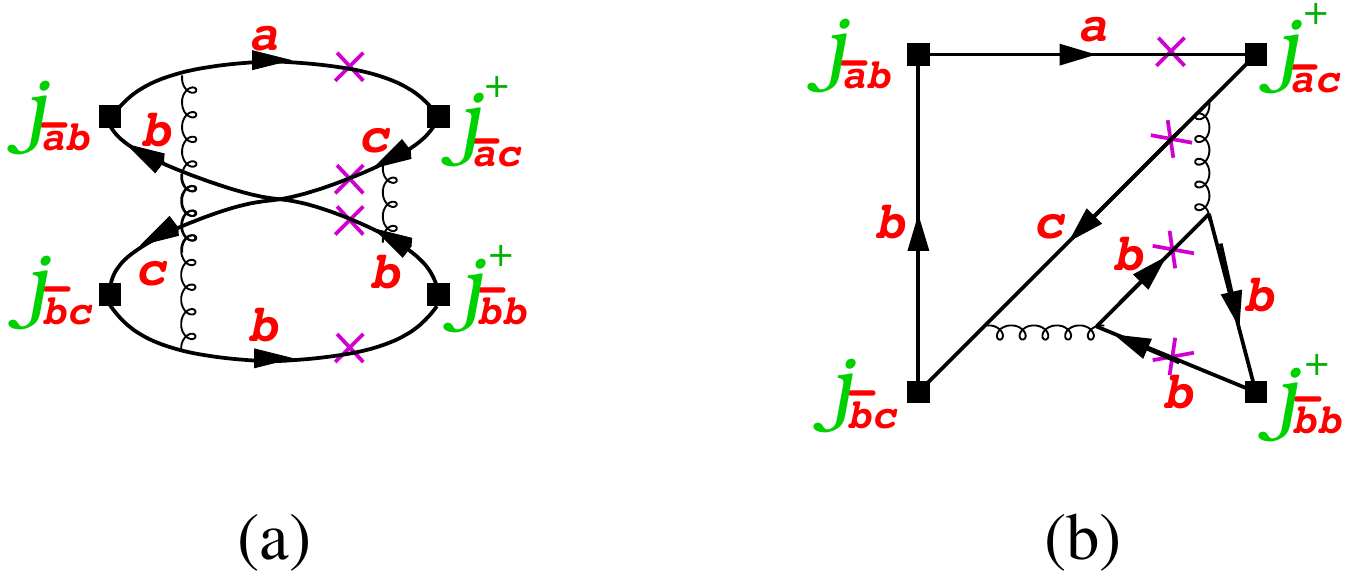} \caption{Representative sample of lowest-order tetraquark-phile perturbative contributions to (top row) flavour-preserving correlation functions (\ref{cfp}) or (bottom row) flavour-reordering correlation functions~(\ref{cfr}). In each of the contributions depicted, curly lines represent the (at lowest relevant order) pair of internally exchanged gluons and the foursome of purple crosses indicates a possible location of any four-quark~cut.}\label{F12ab}\end{figure}

The application of the suggested simple general criterion to the perturbative expansions of these correlation functions demonstrates \cite{TQC}: for all flavour-cryptoexotic tetraquarks (\ref{3F}), each tetraquark-phile contribution needs at least two internal gluon exchanges of suitable~topology. Consequently, tetraquark-phile contributions cannot emerge before the next-to-next-to-lowest perturbative order. Generic examples of the second-order contributions can be found in Fig.~\ref{F12ab}.

If implemented in analyses of exotic hadron states, multiquark-phile improvements of any adopted framework sharpen one's analytic blades by zooming in on the true multiquark nature of one's bound states of interest. For the bound-state approach provided by the QCD sum-rule formalism \cite{QSRa,QSRb}, the tetraquark-adequate modifications have been sketched in Refs.~\cite{TAa,TAb}.

\section{Multitude of qualitative results and predictions from large-\boldmath{$N_{\rm c}$} QCD}\label{Nc}Essentially \textsl{qualitative} insights may be gained by generalizing the gauge group SU(3) of~QCD to some special unitary Lie group of degree $N_{\rm c}$, SU($N_{\rm c}$), and enabling the number $N_{\rm c}$ of \textsl{colour} degrees of freedom to increase without bound, $N_{\rm c}\to\infty$, accompanied by a matching decrease to zero of the strong coupling $g_{\rm s}$, or the strong fine-structure coupling $\alpha_{\rm s}$, at a proper~pace~\cite{Nc}:\begin{align*}g_{\rm s}\propto\frac{1}{\sqrt{N_{\rm c}}}&=O(N_{\rm c}^{-1/2})\xrightarrow[N_{\rm c}\to\infty]{}0\\&\,\Updownarrow\hspace{21.652ex}\Longrightarrow\qquad N_{\rm c}\,g_{\rm s}^2\propto N_{\rm c}\,\alpha_{\rm s}=O(N_{\rm c}^0)=\mbox{const}\ .\\\alpha_{\rm s}\equiv\frac{g_{\rm s}^2}{4\pi}\propto\frac{1}{N_{\rm c}}&=O(N_{\rm c}^{-1})\xrightarrow[N_{\rm c}\to\infty]{}0\end{align*}However, extrapolating the SU(3) quark representation to gauge groups SU($N_{\rm c}$) is, fortunately or unfortunately, far from unique: The by far most common choice is to retain all quarks in the fundamental SU($N_{\rm c}$) representation, of dimension $N_{\rm c}$. However, the $\frac{1}{2}\,N_{\rm c}\,(N_{\rm c}-1)$-dimensional antisymmetric ${\rm SU}(N_{\rm c})$ representation constitutes another option for the assignment of quarks.

Within the wide realm of exotic hadrons, the \textsl{decays} of the tetraquark states of the form~(\ref{fce}) and their allowed \textsl{mixings} with flavour-compatible ordinary mesons (\ref{cm}) are of utmost interest.

\subsection{Large-\boldmath{$N_{\rm c}$} behaviour of decay width of cryptoexotic tetraquarks: upper bounds}For every multiquark hadron, on an equal footing with its mass its total \textsl{decay width} $\Gamma$ belongs to the most crucial characteristics of this hadron state. The coupling strength of a tetraquark $T$ to \textsl{pairs} of conventional mesons $M$, encoded in a related transition amplitude $A(T\longleftrightarrow M\,M)$, can be inferred from the respective tetraquark-pole contribution to correlation functions of the sort (\ref{4}). For all cryptoexotic tetraquark mesons in the current focus, the large-$N_{\rm c}$ dependences of both categories, (\ref{cfp}) and (\ref{cfr}), of \textsl{tetraquark-phile} correlation functions prove to be the~same:\footnote{In contrast to these observations, for tetraquark mesons that involve four mutually unequal quark flavours the two possible sorts of leading-order tetraquark-phile contributions to the related correlation functions differ by one order of $N_{\rm c}$ \cite{TQC}; consistency then necessitates the \textsl{pairwise} appearance \cite{TQC} of such genuinely flavour-exotic tetraquark~mesons.}\begin{equation}\left\langle{\rm T}\!\left(j(y)\,j(y')\,j^\dag(x)\,j^\dag(x')\right)\right\rangle_\textrm{tetraquark-phile}=O(N_{\rm c}^2\,\alpha_{\rm s}^2)=O(N_{\rm c}^0)\ .\label{CP}\end{equation}Taking into account the standard \cite{Nf} $N_{\rm c}$ dependence of \textsl{conventional}-meson decay constants,\begin{equation*}f_{M_{\overline ab}}\equiv\langle0|j_{\overline ab}|M_{\overline ab}\rangle\propto\sqrt{N_{\rm c}}=O(N_{\rm c}^{1/2})\ ,\end{equation*}implies for the (potential) cryptoexotic-tetraquark--two-ordinary-meson transition amplitudes\begin{equation}\underbrace{A(T\longleftrightarrow M_{\overline ab}\,M_{\overline bc})=O(N_{\rm c}^{-1})\stackrel{\mbox{$N_{\rm c}$ order}}{\mbox{\boldmath$\equiv$}}A(T\longleftrightarrow M_{\overline ac}\,M_{\overline bb})=O(N_{\rm c}^{-1})}_{\mbox{$\Longrightarrow\qquad\Gamma(T)=O(N_{\rm c}^{-2})$\ .}}\label{cpo}\end{equation}

If the decays of flavour-cryptoexotic tetraquarks (\ref{fce}) are dominated by any decays into \textsl{two} ordinary mesons (\ref{cm}), the squares of the transition amplitudes (\ref{cpo}) govern the \textsl{total} decay~widths $\Gamma(T)$ of any flavour-cryptoexotic tetraquarks (\ref{fce}), whence the latter are of the order $O(N_{\rm c}^{-2})$~\cite{TQC}.

This prediction, deduced by adopting exclusively tetraquark-phile contributions, however, deviates from the results of similar earlier studies, \textsl{cf.}~Table~\ref{W}. The discrepancies can be traced back \cite{WL:MT2} to either a retainment \cite{KP} of contributions to correlation functions that, evidently,~are not tetraquark-phile from the point of view of Ref.~\cite{TQC}, or a neglect \cite{MPR} of flavour-rearranging contributions to correlation functions exemplified by the graph (b) in the bottom row of Fig.~\ref{F12ab}.

\begin{table}[ht]\caption{Large-$N_{\rm c}$ QCD predictions \cite{LMSurp} of \textsl{upper bounds} on the decrease of the total decay widths $\Gamma(T)$ of the \textsl{flavour-cryptoexotic} tetraquarks (\ref{fce}) with increasing number $N_{\rm c}$ of the colour degrees of freedom.\label{W}}\begin{center}\begin{tabular}{lcr}\toprule\textbf{Author Collective}&\textbf{$\hspace{2.39022ex}$Large-$N_{\rm c}$ Total Decay Width $\Gamma(T)$$\hspace{2.39022ex}$}&\textbf{References}\\\midrule Knecht and Peris&$O(1/N_{\rm c})$&\cite{KP}\\Maiani, Polosa and Riquer&$O(1/N_{\rm c}^3)$&\cite{MPR}\\Lucha, Melikhov and Sazdjian&$O(1/N_{\rm c}^2)$&\cite{TQC}\\\bottomrule\end{tabular}\end{center}\end{table}

Needless to stress, the tetraquark mesons need not betray their actual existence necessarily already at the \textsl{largest} tetraquark-phile order. Consequently, in a strict sense all large-$N_{\rm c}$ claims of the above kind can provide just an upper bound on the respective factual large-$N_{\rm c}$ behaviour of tetraquark decay widths. Anyway, in the large-$N_{\rm c}$ limit the flavour-cryptoexotic tetraquarks are narrow states: for increasing $N_{\rm c}$, their total decay widths $\Gamma(T)$ decrease and this even~faster than the ``well-known'' decrease \cite{Nf} of the decay widths $\Gamma(M)$ of the conventional mesons~$M$,\begin{equation*}\Gamma(T)=O(N_{\rm c}^{-2})\stackrel{\mbox{$N_{\rm c}$ order}}{\mbox{\boldmath$<$}}\Gamma(M)\propto\frac{1}{N_{\rm c}}=O(N_{\rm c}^{-1})\xrightarrow[N_{\rm c}\to\infty]{}0\ .\end{equation*}Such kind of narrowness definitely promotes the empirical detectability of (unstable) hadrons: in comparison with the \textsl{mass} of any such hadron, its total \textsl{decay width} should not be too broad.

\subsection{Mixing of cryptoexotic tetraquark and same-open-flavour conventional meson}In contrast to all manifestly flavour-exotic tetraquarks featuring four open quark flavours \cite{WL:MT1}, for every flavour-cryptoexotic tetraquark state the, at least, potential \textsl{mixing} with conventional mesons of compatible quantum numbers should be anticipated. Inspecting the contribution of a single intermediate mixing of a flavour-cryptoexotic tetraquark (\ref{3F}) with ``flavour-congruent'' conventional mesons $M_{\overline ac}$ to a correlation function (\ref{CP}) straightforwardly leads to the constraint\begin{equation*}\underbrace{A(T\longleftrightarrow M_{\overline ab}\,M_{\overline bc})}_{\mbox{$O(N_{\rm c}^{-1})$}}\,g_{TM_{\overline ac}}\,\underbrace{A(M_{\overline ac}\longleftrightarrow M_{\overline ab}\,M_{\overline bc})}_{\mbox{$O(N_{\rm c}^{-1/2})$}}=O(N_{\rm c}^{-2})\end{equation*}on the products of a tetraquark--two-ordinary-meson transition amplitude $A(T\longleftrightarrow M_{\overline ab}\,M_{\overline bc})$, a tetraquark--conventional-meson mixing strength $g_{TM_{\overline ac}}$ and a three-ordinary-meson coupling $A(M_{\overline ac}\longleftrightarrow M_{\overline ab}\,M_{\overline bc})$, with any latter amplitude established \cite{Nf} to decrease, for large $N_{\rm c}$,~like\begin{equation*}A(M_{\overline ac}\longleftrightarrow M_{\overline ab}\,M_{\overline bc})\propto\frac{1}{\sqrt{N_{\rm c}}}=O(N_{\rm c}^{-1/2})\xrightarrow[N_{\rm c}\to\infty]{}0\ .\end{equation*}So, the tetraquark-phile findings (\ref{CP}) yield for the large-$N_{\rm c}$ behaviour of the mixing strength~\cite{TQC}\begin{equation*}g_{TM_{\overline ac}}\propto\frac{1}{\sqrt{N_{\rm c}}}=O(N_{\rm c}^{-1/2})\xrightarrow[N_{\rm c}\to\infty]{}0\ .\end{equation*}This observation is by no means surprising: comparing the mixing case with the case of a pure tetraquark-pole contribution betrays just a replacement of the tetraquark--two-ordinary-meson amplitude by the product of a mixing strength and some three-conventional-meson amplitude. The large-$N_{\rm c}$ decrease of the interchanged intermediate expressions is, qualitatively, identical:\begin{align*}A(T\longleftrightarrow M_{\overline ab}\,M_{\overline bc})&=O(N_{\rm c}^{-1})\\&\Updownarrow\\g_{TM_{\overline ac}}\,A(M_{\overline ac}\longleftrightarrow M_{\overline ab}\,M_{\overline bc})&=O(N_{\rm c}^{-1})\ .\end{align*}Consequently, the mixing of a flavour-cryptoexotic tetraquark with conventional mesons does not possess the capability to alter the large-$N_{\rm c}$ behaviour of this tetraquark's total decay width.

\section{Brief summary of minimally surprising outcomes and conclusions}Almost by definition, the total quark content of each \textsl{multiquark} state can be decomposed~into (more than one) colour-singlet bound states each of which involves a smaller number of quark constituents, and ultimately into a sequence of conventional hadrons. Consequently, as is well known \cite{SW}, theoretical analyses of multiquarks have to face the risk of being contaminated or corrupted by contributions providing no information about the multiquark hadrons of interest.

In view of these obstacles, a criterion has been formulated \cite{TQC} that promises to allow for an (unambiguous) identification of precisely all those contributions to one's preferred~framework that bear some relationship to the tetraquark analyzed and for a rigorous rejection of all others.

Application of this \textsl{selection tool} to the drastic simplification offered by the generalization of quantum chromodynamics to the set of quantum field theories characterized by the increase beyond bound of their number of colour degrees of freedom, accompanied by the (concurrent) decrease of the strong coupling strength, to the (from an experimental point of view) currently most densely populated multiquark subdivision of all flavour-cryptoexotic tetraquarks proves:\begin{itemize}\item In this (delicate) limit, the total decay widths of all flavour-cryptoexotic tetraquark states~(\ref{3F}) vanish in proportion to the inverse of the square of the number of colour degrees of freedom.\item This decrease to zero proceeds parametrically even faster than that of a conventional meson.\clearpage\item This decrease is not affected by the potentially occurring mixing of the flavour-cryptoexotic tetraquark state with a conventional meson exhibiting \textsl{identical} open quark-flavour content.\end{itemize}Beyond doubt, the considerations of Sects.~\ref{4.cor} or \ref{Nc} and the derived implications, exemplified by the flavour-cryptoexotic tetraquarks~(\ref{3F}), ought to be generalized to the entire set of multiquark hadrons and eventually be reflected in future quantitative predictions of multiquark properties.

\section*{Acknowledgements}W.~L.\ thanks both D.~Melikhov and H.~Sazdjian for pleasant collaboration on the above topics.

\end{document}